\documentclass[12pt]{article}
\usepackage{amsmath}
\usepackage{amssymb}
\usepackage{geometry}
\usepackage{hyperref}
\usepackage{graphicx}
\usepackage{cite}
\usepackage{tabularx} % Include this in your preamble
\usepackage{booktabs} % For improved table aesthetics
\usepackage{siunitx}  % For formatting numbers
\usepackage{float}
\usepackage{siunitx} % For formatting numbers
\usepackage{booktabs} % For improved table aesthetics
\usepackage{rotating} % For rotating the table horizontally

\usepackage{array}

\begin{document}

\title{Universal Dynamics of Financial Bubbles in Isolated Markets: Evidence from the Iranian Stock Market}
\author{%
{\large Ali Hosseinzadeh}\\[4pt]
{\small Department of Physics, Shahid Beheshti University}\\
{\small Evin, Tehran 1983969411, Iran}\\
{\small \texttt{ali\_hosseinzadeh@sbu.ac.ir}}%
}
\date{} % removes date

\maketitle

\begin{abstract}
Speculative bubbles exhibit common statistical signatures across many financial markets, suggesting the presence of universal underlying mechanisms. We test this hypothesis in the Iranian stock market, an economy that is highly isolated, subject to capital controls, and largely inaccessible to foreign investors. Using the Log-Periodic Power Law Singularity (LPPLS) model, we analyze two major bubble episodes in 2020 and 2023. The estimated critical exponents ($\beta \approx 0.46$ and $\beta \approx 0.20$) fall within the empirical ranges documented for canonical historical bubbles such as the 1929 DJIA crash and the 2000 Nasdaq episode. The Tehran Stock Exchange displays clear LPPLS hallmarks: faster-than-exponential price acceleration, log-periodic corrections, and stable estimates of the critical time horizon. These results indicate that endogenous herding, imitation, and positive-feedback dynamics---rather than exogenous shocks---play a dominant role even in politically and economically isolated markets. By showing that an emerging and semi-closed financial system conforms to the same dynamical patterns observed in global markets, this paper provides new empirical support for the universality of bubble dynamics. To the best of our knowledge, it also presents the first systematic LPPLS analysis of bubbles in the Tehran Stock Exchange. The findings highlight the usefulness of LPPLS-based diagnostic tools for monitoring systemic risk in emerging or restricted economies.
\end{abstract}

\section{Introduction}

Speculative bubbles and market crashes are widely observed phenomena in financial systems around the world. Over the past two decades, numerous studies have suggested that such events may share common underlying dynamics, independent of geography, market structure, or regulatory environment. In particular, the Log-Periodic Power Law Singularity (LPPLS) model—rooted in statistical physics and complex systems theory—has been successfully applied to a range of financial bubbles, from the 1929 Wall Street crash to the 2000 Nasdaq collapse~\cite{Sornette2003, Johansen1999, Sornette2014}. These studies have revealed recurring features such as faster-than-exponential growth, log-periodic oscillations, and identifiable precursors to critical transitions.

While the LPPLS model has been successfully applied to various well-integrated and highly liquid markets, its validity in structurally isolated or emerging financial systems remains underexplored. Most empirical studies have focused on bubbles in advanced economies, where data accessibility, institutional stability, and high-frequency trading behaviors support robust model calibration~\cite{Sornette2003, Zhou2006, Nielsen2024}. 

The Iranian stock market, in contrast, presents a unique opportunity to test whether the endogenous mechanisms captured by LPPLS—such as herding, reflexivity, and log-periodic behavior—still manifest in a market characterized by international sanctions, restricted capital flows, and limited foreign investor access~\cite{OddLotsPodcast}. 

This raises a broader theoretical question: To what extent are the dynamics of financial bubbles universal, and do they hold even in markets that are isolated from the global financial system?

While the LPPLS model has been extensively validated in well-integrated, liquid markets, its applicability in emerging or structurally isolated markets—such as Iran—remains an open empirical and theoretical question.
Prior studies have primarily focused on developed economies, where institutional maturity, transparency, and data availability support robust calibration~\cite{Sornette2003,Zhou2006,Farmer2005}. In contrast, the Tehran Stock Exchange (TSE) operates under international sanctions, limited foreign participation, and a retail-driven structure—conditions rarely tested under the LPPLS framework.

Addressing this gap not only advances the empirical frontier of LPPLS applications but also contributes to a larger conversation in econophysics and complex systems theory: whether endogenous mechanisms like herding, positive feedback, and reflexivity exhibit universal behavior across financial systems regardless of geopolitical or structural isolation~\cite{Yakovenko2009, Gabaix2003}.

This paper builds upon this body of work by investigating whether these universal features of bubble dynamics also appear in an economically and politically isolated market: the Tehran Stock Exchange (TSE). We propose that financial instabilities, such as those observed during the 2020 and 2023 crashes in Iran, conform to generalizable principles of collective behavior and criticality—even in the absence of global capital flows or mature institutional frameworks. This hypothesis aligns with recent perspectives in econophysics, which emphasize the role of endogenous feedback mechanisms like herding, imitation, and self-organization in driving speculative dynamics~\cite{Farmer2005, Yakovenko2009}.

\vspace{1em} % ---- preserve your original paragraph flow below ----

In the podcast episode \textit{What's Been Happening With the Iranian Stock Market} by Bloomberg's Odd Lots, Maciej Wojtal, a London-based fund manager specializing in Iranian stocks, describes the Tehran Stock Exchange as ``one of the world's most unfamiliar markets''~\cite{OddLotsPodcast}. This unfamiliarity arises primarily due to international sanctions that limit access to accurate information and restrict foreign investment in Iranian stocks. Despite Iran being a large middle-income country with significant economic potential, its stock market remains largely isolated from global financial systems, rendering it a niche area for specialized investors. The main contribution of this paper is to provide the first systematic application of the LPPLS model to speculative bubbles in the Tehran Stock Exchange. By analyzing two major episodes—the 2020 crash and the 2023 bubble—we show that the estimated LPPLS exponents fall within the empirical ranges reported for canonical bubbles in major international markets. This demonstrates that bubble dynamics in an isolated, sanction-constrained economy share the same universal signatures as those in more integrated systems. In addition, we complement the parametric LPPLS analysis with a non-parametric log-periodic detection procedure and contextual “stories” that link the bubble phases to macroeconomic conditions and policy interventions in Iran.

This paper aims to demystify the Tehran Stock Exchange by analyzing the bubble dynamics associated with the 2020 crash and the 2023 bubble. We utilize the Log-Periodic Power Law Singularity (LPPLS) model to demonstrate that, despite the market’s unfamiliarity, it exhibits patterns similar to other global stock markets. Our findings suggest that, using only data available before each crash, there was scope to anticipate these bubbles in the Tehran stock market, akin to the predictability documented in more familiar markets.

The 2020 crash had a profound impact on the Tehran Stock Exchange, with the total market index plummeting by approximately 42\% (Figure 1). This event stands as the most significant crash in the market's history. While some experts attribute the crash to political issues, our analysis indicates that positive feedback mechanisms were the primary drivers. For example, the Persian Gulf Petrochemical Industries Corporation (PGPIC), the largest company on the Tehran Stock Exchange, saw its valuation peak at \$22.4~billion during the bubble, only to decline to around \$10~billion currently. Similarly, the Social Security Investment Company (SSIC), also known by its Persian acronym SHASTA, experienced a peak valuation of approximately \$24~billion, which has since decreased to around \$3~billion.

In the following sections, we will present the results of our LPPLS analysis, shedding light on the bubble characteristics of the Tehran stock market during this period and reinforcing the potential for bubble predictability even in less familiar markets.

\begin{figure}
    \centering
    \includegraphics[width=0.91\linewidth]{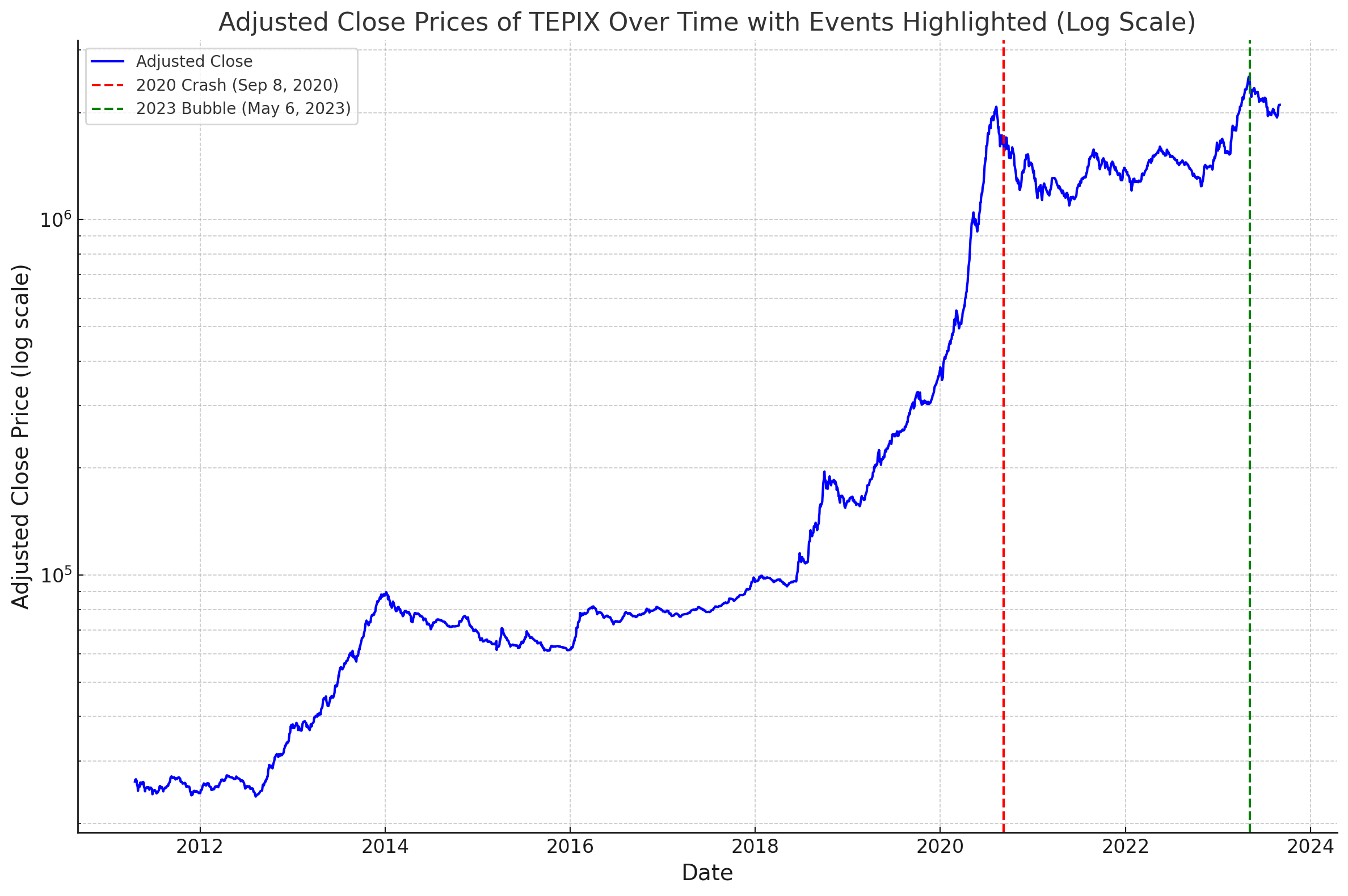}
    \caption{The 2020 crash (red dashed line) and the 2023 bubble (green dashed line)}
\end{figure}

\section{Review of the LPPLS Model and Positive Feedback Mechanisms}

\subsection{Overview of the LPPLS Model}

The Log-Periodic Power Law Singularity (LPPLS) model is a mathematical framework used to describe and predict the behavior of asset prices during financial bubbles and crashes~\cite{Sornette2003,Sornette2017}. Developed by Didier Sornette and his collaborators, the LPPLS model captures the characteristic super-exponential growth of asset prices leading up to a critical point, often followed by a market correction or crash. The model incorporates both the accelerating growth of prices and the oscillatory behavior caused by investor herding and positive feedback mechanisms.

\subsection{Success and Applications of the LPPLS Model}

The LPPLS has been successfully applied to a wide range of historical bubbles, including the 1929 Wall Street crash, the 1987 Black Monday, the 2000 dot-com bubble, the 2006–2008 housing market crash, and even recent phenomena such as the 2020 COVID-induced crash in the S\&P 500~\cite{Sornette2003, Johansen1999, Sornette2017,  Zhang2022}.

The LPPLS model captures two key features of financial bubbles: faster-than-exponential price growth driven by positive feedback mechanisms (such as herding and momentum trading), and log-periodic oscillations that arise from discrete scale invariance and hierarchical structure in trader interactions. These oscillations often manifest as volatility clustering and can serve as early-warning signals of regime shifts. Numerous empirical studies have shown that the model not only fits historical data well, but also provides ex-ante predictive power when carefully calibrated~\cite{Zhou2006, Sornette2012}.

Its success lies in its capacity to model endogenous market dynamics using principles from statistical physics and complex systems, bridging economic theory with mathematical rigor. In the context of this study, applying the LPPLS model to the Iranian market provides a robust framework for comparing the universality of bubble dynamics across structurally different financial systems.

\subsection{Mathematical Formulation}

The LPPLS model posits that the logarithm of an asset's price $p(t)$ can be described by a function that includes a power law term and a log-periodic oscillation term. The general form of the LPPLS equation is:

\begin{equation}
\ln p(t) = A + B (t_c - t)^\beta + C (t_c - t)^\beta \cos\left[ \omega \ln (t_c - t) + \phi \right],
\label{eq1}
\end{equation}

where:

\begin{itemize}
    \item $\ln p(t)$ is the natural logarithm of the asset price at time $t$.
    \item $A$ is a constant representing the baseline level of $\ln p(t)$.
    \item $B$ and $C$ are amplitudes of the power law growth and the oscillatory terms, respectively.
    \item $t_c$ is the critical time at which the bubble is expected to burst or the market is expected to crash.
    \item $\beta$ is the exponent of the power law term, typically constrained between 0 and 1 ($0 < \beta < 1$).
    \item $\omega$ is the angular frequency of the logarithmic periodic oscillations.
    \item $\phi$ is a phase shift that adjusts the starting point of the oscillations.
\end{itemize}

\subsubsection{Interpretation of the Parameters}

\begin{itemize}
    \item \textbf{Power Law Term ($B (t_c - t)^\beta$)}: This term captures the accelerating growth of prices as they approach the critical time $t_c$. The exponent $\beta$ determines the rate of this acceleration.
    \item \textbf{Log-Periodic Oscillation Term ($C (t_c - t)^\beta \cos\left[ \omega \ln (t_c - t) + \phi \right]$)}: This term accounts for the oscillatory behavior observed in asset prices due to collective investor behaviors like herding. The frequency $\omega$ and phase $\phi$ characterize the pattern of these oscillations.
\end{itemize}

\subsection{Positive Feedback and Bubble Formation}

Positive feedback mechanisms are central to the formation and growth of financial bubbles~\cite{Sornette2003}. In the context of the LPPLS model, positive feedback refers to the self-reinforcing process where rising asset prices attract more investors, further driving up prices. This phenomenon is often fueled by psychological factors such as:

\begin{itemize}
    \item Herding behavior refers to the tendency of investors to follow the actions of the majority—buying assets simply because others are buying, regardless of underlying fundamentals~\cite{Bikhchandani1992, Banerjee1992, Devenow1996}.

    \item \textbf{Expectation of Future Price Increases}: The belief that prices will continue to rise motivates more buying, creating a feedback loop.
    \item \textbf{Speculative Investment}: Investors purchase assets with the intent to sell at higher prices in the short term, contributing to price inflation.
\end{itemize}

As a result of positive feedback, asset prices deviate significantly from their intrinsic values, leading to unsustainable growth patterns that the LPPLS model aims to capture. The model's ability to characterize the super-exponential growth and oscillatory behavior of prices makes it a valuable tool for predicting the timing of market corrections or crashes.

\subsection{Application to the Tehran Stock Market}

Applying the LPPLS model to the Tehran Stock Exchange allows for the analysis of bubble dynamics in an unfamiliar and largely isolated market. Despite the lack of widespread international engagement due to sanctions, the Tehran stock market exhibits patterns consistent with other global markets experiencing bubbles. The positive feedback mechanisms observed---such as rapid price increases attracting more investors---mirror those in more familiar markets.

By fitting the LPPLS model to the Tehran stock market data leading up to the 2020 crash and the 2023 bubble, we aim to demonstrate that:

\begin{itemize}
    \item The market experienced super-exponential growth indicative of a financial bubble.
    \item When calibrated on data up to several weeks before the crash, the model can predict a critical time $t_c$ that aligns closely with the eventual crash date.
    \item Positive feedback mechanisms were significant contributors to the bubble's formation and eventual burst.
\end{itemize}

In the next section, we will delve into the empirical results of our LPPLS analysis on the Tehran stock market, highlighting \textbf{how the model's parameters reflect the market's dynamics during the bubble period.}

\section{Market Structure and Performance of the Tehran Stock Exchange}

\subsection{Market Structure of TSE}

The Tehran Stock Exchange (TSE) comprises multiple segments catering to different financial instruments and investor types. The largest and most active segment is the \textit{Equities Market}, where publicly traded companies issue and trade shares. This market consists of a primary segment for new stock issuance and a secondary segment for trading existing shares. Companies are listed on either the \textit{First Market (Main Board)}, which includes well-established firms meeting stringent listing requirements, or the \textit{Second Market}, where smaller and less liquid companies operate.

Alongside equities, the \textit{Debt Market} facilitates investment in government bonds, Sukuk (Islamic bonds), and corporate debt securities, providing stability for risk-averse investors. The \textit{Derivatives Market} offers options and futures contracts, enabling investors to hedge against price fluctuations. Furthermore, \textit{Exchange-Traded Funds (ETFs)} have gained prominence, with funds tracking both equity indices and fixed-income securities. Finally, the \textit{Professional Investment Market} serves institutional investors and specialized financial products, including structured funds.

\subsection{Trading System and Rules}

TSE operates under a structured regulatory framework that dictates market behavior. Trading is conducted from Saturday to Wednesday, between 9:00 AM and 12:30 PM local time. To mitigate excessive volatility, daily price fluctuations are restricted to \textpm 5\%. When a stock reaches the upper limit and the demand for buying exceeds the available supply, a buy queue  is formed. Conversely, when a stock hits the lower limit and selling pressure dominates, a sell queue  emerges.

The fluctuation limit plays a crucial role in stabilizing prices by preventing excessive volatility within a single trading day, but it also directly contributes to the formation of these buy and sell queues, which can delay price discovery.
Additionally, transactions follow a T+2 settlement cycle, meaning trades are finalized two business days after execution. The T+2 settlement system refers solely to the cash and ownership transfer timing and does not impose a minimum holding period for stocks; investors can sell their shares at any time during the same day if market conditions allow.

\subsection{Key Features and Challenges of TSE}

TSE presents a mix of strengths and challenges. A key advantage is its diverse investment landscape, offering opportunities across equities, bonds, ETFs, and derivatives. Regulatory oversight by the Securities and Exchange Organization (SEO) further enhances market integrity and investor confidence. Additionally, market capitalization has expanded significantly in recent years, signaling the growing importance of TSE within the region.

Despite these strengths, challenges persist. Foreign investment remains restricted due to economic sanctions, limiting international capital inflows. Liquidity constraints affect certain stocks, making large-volume trades difficult. Moreover, the fixed daily price fluctuation limits can sometimes result in artificial pricing distortions and sharp market swings.

\section{Tehran Stock Exchange (TSE) Market Overview in 2020}

In order to analyze the Tehran Stock Exchange (TSE) around mid-2020 and detect potential bubble-like behavior, we extract key statistics from official reports\footnote{This section is based on official reports of the TSE, which can be found in \cite{SEOAnnualReports}.}. Below is a comprehensive market overview for that period.

\subsection{Number of Public Companies Traded}
By the end of March 2021, the number of listed companies across different markets was:
\begin{itemize}
    \item \textbf{Tehran Stock Exchange (TSE)}: 369 companies
    \item \textbf{Iran Fara Bourse (IFB)}: 141 companies
    \item \textbf{Base Market (OTC)}: 174 companies
    \item \textbf{Total listed companies}: 684 entities
\end{itemize}

\subsection{Market Capitalization}
The total market capitalization of the Tehran Stock Exchange in March 2021 was 72,418,800 billion IRR (approx. 359 billion \$), marking a 177\% increase compared to the previous year.

Breakdown by financial instrument type:
\begin{table}[h]
    \centering
    \begin{tabular}{|l|c|}
        \hline
        \textbf{Financial Instrument} & \textbf{Market Share (\%)} \\
        \hline
        Equities (Stock Market) & 23 \\
        Debt Market (Fixed-Income Securities) & 58 \\
        ETFs & 8 \\
        Physical Commodities \& Energy Exchange & 11 \\
        \hline
    \end{tabular}
    \caption{Market Capitalization Breakdown in 2020}
    \label{tab:market_cap}
\end{table}

\subsection{Trading Frequency and Volume}

For the Iranian fiscal year 1399 (20 March 2020--20 March 2021), official statistics from
the Securities and Exchange Organization (SEO) report that the total number of stock trades
across the Tehran Stock Exchange and Iran Fara Bourse reached 559{,}597{,}153 transactions.
Over the same period, aggregate trading volume amounted to 2{,}825{,}997{,}294 thousand
shares (roughly $2.83\times 10^{12}$ shares), with a total traded value of 33{,}882{,}952~billion
IRR. The reported daily averages over 1399 were 27{,}251{,}350 trades, 138{,}499{,}921 thousand
shares (about $1.38\times 10^{11}$ shares), and 1{,}642{,}621~billion IRR in trading value.\footnote{All
figures in this subsection are taken from the SEO's end-of-year report for 1399.}

\subsection{Sectoral Breakdown \& Key Industries}
\textbf{Top Traded Sectors by Market Share:}
\begin{table}[h]
    \centering
    \begin{tabular}{|l|c|}
        \hline
        \textbf{Industry} & \textbf{Trading Value (Billion IRR)} \\
        \hline
        Oil, Gas, and Petroleum Products & 203,985 \\
        Basic Metals (Steel, Copper, Aluminum) & 139,466 \\
        Banking and Financial Services & 103,004 \\
        Chemical Products & 149,670 \\
        \hline
    \end{tabular}
    \caption{Top Traded Sectors in 1399}
    \label{tab:sectors}
\end{table}

\subsection{Trading Statistics in August 2020}
\begin{table}[h]
    \centering
    \begin{tabular}{|l|c|c|c|}
        \hline
        \textbf{Exchange} & \textbf{Stock Trading Value (Billion IRR)} & \textbf{MoM Change (\%)} & \textbf{YoY Change (\%)} \\
        \hline
        TSE & 3,343,361 & -10.7 & +1561.5 \\
        IFB & 1,106,645 & -19.1 & +735.2 \\
        \hline
    \end{tabular}
    \caption{Trading Statistics in August 2020}
    \label{tab:trading_august}
\end{table}

\textbf{Observations:}
\begin{itemize}
    \item Stock trading value dropped \textbf{10.7\%} in TSE and \textbf{19.1\%} in IFB, confirming reduced liquidity.
    \item Massive \textbf{1561\% YoY increase} before the crash suggests bubble-like conditions.
\end{itemize}

\subsection{Bubble Genesis: The “Stories” Behind the 2020 Crashes }

The 2020 Tehran Stock Exchange bubble was catalyzed by a multifaceted “story” built on fiscal desperation, state-led privatization, and a mass behavioral shift in investment practices. 

Amid reimposed U.S. sanctions following the collapse of the Joint Comprehensive Plan of Action (JCPOA), the Iranian government actively encouraged investment in the stock market as a means of compensating for its fiscal deficits. A wave of state-owned enterprise IPOs—most notably including major players like \textit{Shasta}—was promoted as a national wealth-building initiative. Retail investors, many of them first-time participants and explicitly incentivized by state messaging, entered the market en masse.

At the same time, concurrent hyperinflation and the severe devaluation of the Iranian rial  positioned equities as a perceived hedge against inflation. These economic pressures, combined with speculative momentum and extensive government propaganda, pushed valuations far beyond economic fundamentals.

The TEDPIX index ultimately rallied over 190\% during 2020, before the eventual crash reflected the unsustainability of this speculative boom in the face of deteriorating macroeconomic and geopolitical conditions. This combination of sanctions, inflationary pressure, and state-led promotion of equity investment created a fertile environment for strong positive feedback and herding into stocks, consistent with the LPPLS bubble regime identified in our empirical analysis.

\subsection{Evidence of a Bubble Formation }
The TSE saw an extraordinary bull run from late 2019 to mid-2020, followed by a sharp correction in the second half of 1399 (Iranian calendar).

\textbf{Key signs of a bubble:}
\begin{itemize}
    \item \textbf{Massive market capitalization growth} (177\% YoY increase).
    \item \textbf{Exponential price increases across major indices}.
    \item \textbf{P/E ratio surged above historical averages}, indicating speculative investments.
    \item \textbf{Daily trade volumes skyrocketed}, driven by retail participation.
\end{itemize}

\begin{figure}[h]
    \centering
    \includegraphics[width=1\linewidth]{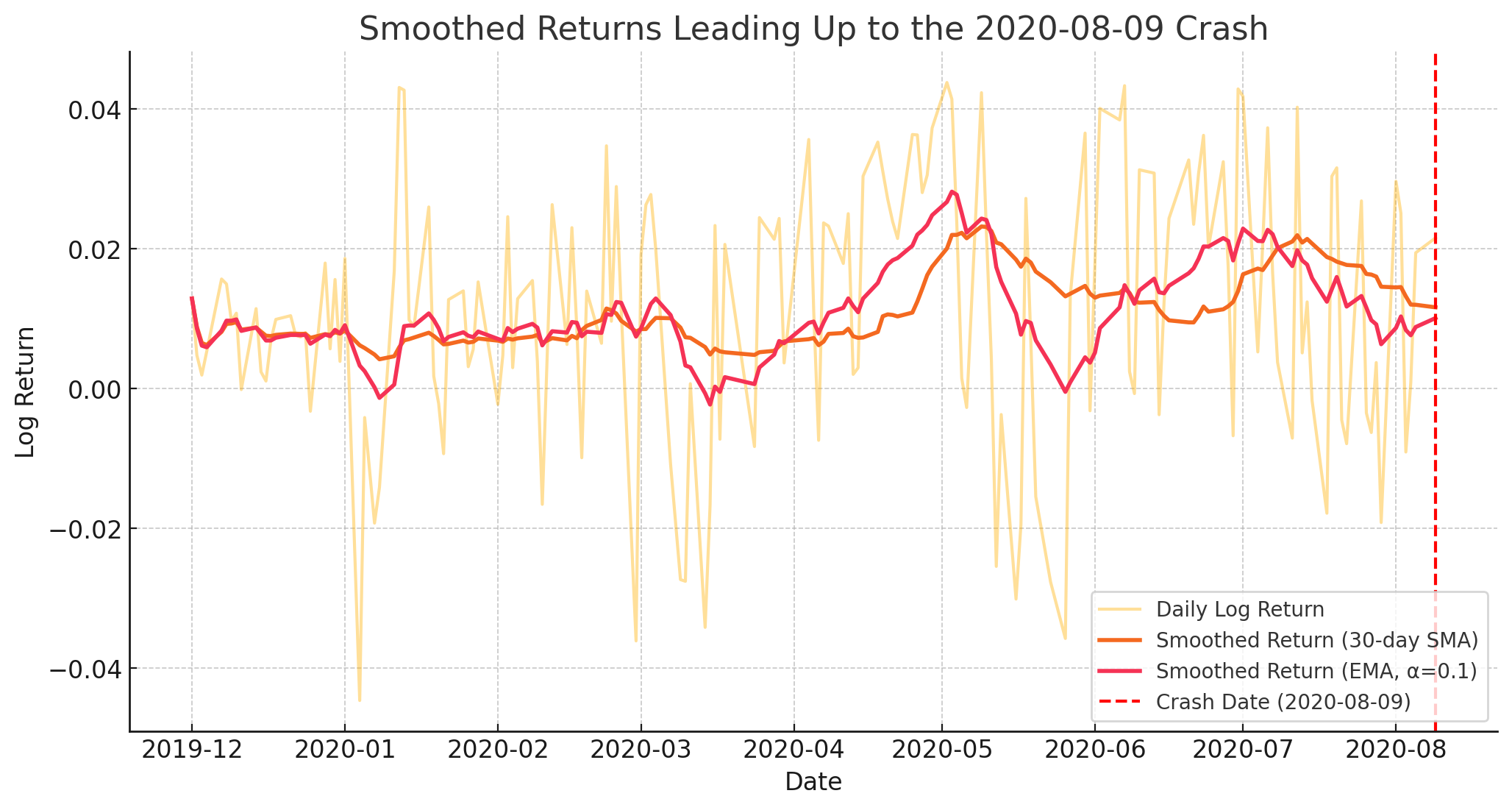}
    \caption{Smoothed Returns Leading Up To The 2020-08-09 Crash}
    \label{fig2}
\end{figure}

This aligns with the LPPLS model criteria for bubble detection. Also, we plot \ref{fig2} which is Smoothed Returns Leading Up To The 2020-08-09 Crash. Both SMA and EMA show a rising trend in returns from late 2019 until the crash. This suggests super-exponential growth, which is a hallmark of speculative bubbles. In a normal market, returns fluctuate around a mean, but here, they increase progressively, meaning investors are driving up prices faster and faster.

\subsection{Conclusion}
The TSE in mid-2020 exhibited strong indications of a speculative bubble, characterized by high market participation, extreme liquidity influx, and super-exponential price growth. This ultimately resulted in a major market correction.

\section{Analysis of the 2020 Tehran Stock Market Crash: Black Sunday}

In this section, we introduce the empirical framework used in this paper and apply it to the 2020 Tehran Stock Exchange bubble. We first present the data and LPPLS calibration scheme for the Tehran market, then explain how we scan over fitting windows to identify bubble inception dates using a Lagrange-regularized procedure. Finally, we outline the non-parametric Lomb–Scargle and (H, q)-derivative diagnostics used to validate log-periodic structures.

\subsection{Data Collection and Preprocessing}

The data used in this analysis comprises the Adjusted Close prices of the Tehran Stock Exchange (TSE) index from January 2017 to August 2020. Due to the unique characteristics of the TSE, including limited accessibility and less frequent trading days compared to other global markets, careful preprocessing was necessary. Non-trading days and public holidays were accounted for to ensure continuity in the time series data. The data was filtered to create multiple time windows for analysis, allowing us to assess the robustness and consistency of the LPPLS model predictions across different periods leading up to the 2020 crash.

\subsection{LPPLS Model Fitting Procedure}

To evaluate how well the 2020 Tehran stock market crash could be anticipated using only pre-crash data, we fitted the LPPLS model to the Adjusted Close price data over various time intervals.
 The calibration of the LPPLS model was performed using the robust and stable two-step nonlinear estimation scheme proposed by Filimonov and Sornette~\cite{FilimonovSornette2013}. In this method, the nonlinear parameters \( (t_c, \beta, \omega) \) are explored over a defined grid, and for each candidate triplet, the corresponding linear parameters \( (A, B, C_1, C_2) \) are estimated via ordinary least squares. The optimal parameter set is selected as the one minimizing the root-mean-square error (RMSE) between the logarithmic observed prices and the LPPLS fit. This approach ensures robustness against noise and enhances convergence, especially in the presence of pronounced log-periodic components.
 Constraints were applied to the parameters based on theoretical considerations:

\begin{itemize} \item The exponent $\beta$ was constrained within $(0,1)$ to ensure super-exponential growth. \item The angular frequency $\omega$ was set to be positive to capture the oscillatory behavior. \item The critical time $t_c$ was restricted to be after the last data point in each time window to prevent fitting past data with future events. \end{itemize}

To evaluate the predictive power of the model, we fitted\footnote{See Appendix I. }
 it over multiple time windows and examined how the estimated critical time \( t_c \) aligned with the actual crash date. However, this methodology lacks robustness. Therefore, in the following sections, we adopt the approaches proposed in~\cite{Demos2019,Demos2017} to systematically identify the inception of financial bubbles, \textbf{rather than relying on arbitrary time window selection.} In our analysis, \( t_1 \) denotes the \textbf{start date} of the fitting window, representing a candidate point where the financial bubble may have begun, while \( t_2 \) is the \textbf{end date}, typically fixed at the known crash date or shifted earlier for robustness testing. The LPPLS model is calibrated over each window \( [t_1, t_2] \) to evaluate which start date best captures the bubble dynamics. This \( [t_1, t_2] \) notation is a common jargon in LPPLS analysis used to define the calibration window for bubble detection.

\subsection{Different Windows with a Fixed $t_2$ }

First of all, approximately one year before the crash, the LPPLS fitting tends to become more precise~\cite{Sornette2003}. Moreover, we generally expect that the estimated critical time \( t_c \) will be close to, but systematically later than, the actual crash date~\cite{Sornette2003}.

\begin{figure}[h]
    \centering
    \includegraphics[width=1\linewidth]{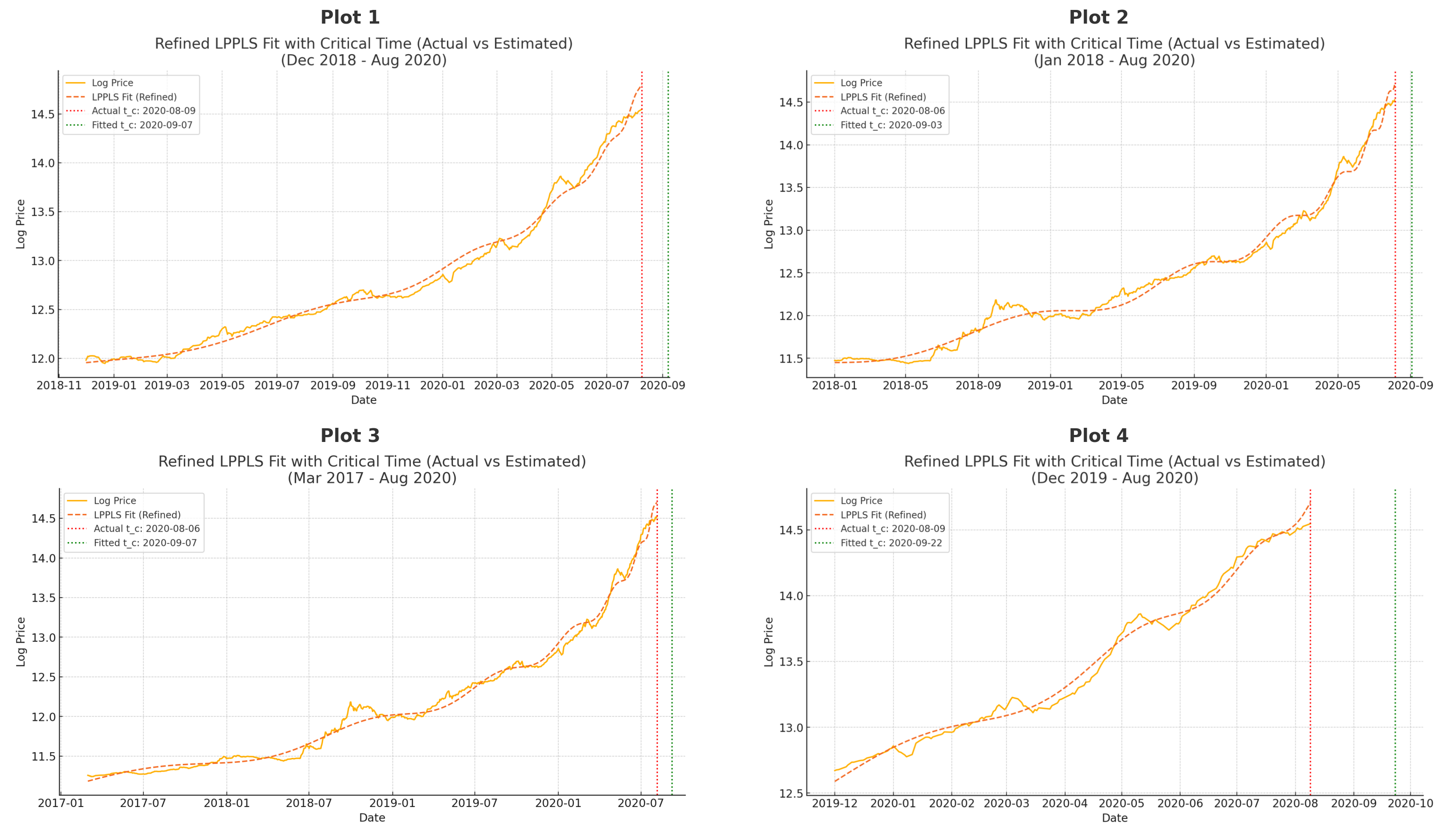}
    \caption{Results for LPPLS Model on different time frames (2020 bubble)}
    \label{fig:3}
\end{figure}

Here, we fixed $t_2$ as the actual crash time to fit the LPPLS model, as shown in Figures~\ref{fig:3} and Table~\ref{tab:lppls_plots_1to4}. Nevertheless, we will show (in section 5.4) that plot 4 in Figure~\ref{fig:3} represents the optimal choice \footnote{This optimized choice is, of course, an approximation, as the parameters vary in a sensitive manner, leading to an uncertainty in the prediction of $t_c$ by approximately several weeks.
}, using Lagrange regularization to further justify this selection. In the next subsection, we also conduct a robustness test by shifting $t_2$ backward to evaluate the stability of the results.

\begin{table}[H]
\centering
\caption{LPPLS Model Parameters for Plots 1–4 (2020 crash)}
\begin{tabular}{lcccc}
\toprule
\textbf{Parameter} & \textbf{Plot 1} & \textbf{Plot 2} & \textbf{Plot 3} & \textbf{Plot 4} \\
\midrule
$A$       & 22.8640  & 22.6180  & 22.9161  & 16.2747 \\
$B$       & -5.7278  & -5.5960  & -5.7556  & -0.2767 \\
$C$       & 0.0288   & 0.0515   & 0.0417   & 0.0077 \\
$\beta$   & 0.1      & 0.1      & 0.1      & 0.4556 \\
$\omega$  & 10.6667  & 11.25    & 11.25    & 9.8889 \\
$\phi$    & -0.8462  & 2.3792   & 2.1545   & 1.6631 \\
$t_c$ (estimated) & 2020-09-07 & 2020-09-03 & 2020-09-07 & 2020-09-22 \\
$t_c$ (days from start) & 646.39 & 976.39   & 1286.27  & 296.90 \\
\bottomrule
\end{tabular}
\label{tab:lppls_plots_1to4}
\end{table}

\subsubsection{Discussion of Results}

These results support the hypothesis that the Tehran stock market bubble was driven by
positive feedback mechanisms, as captured by the LPPLS model. The ability of the model,
when calibrated solely on pre-crash data, to predict a critical time close to the realized crash reinforces its applicability even in less familiar markets like the TSE.

\subsection{Methodology: Bubble Start Detection via Lagrange-Regularized LPPLS} \label{5}

To identify the inception of speculative bubbles preceding major financial crashes, we apply the Log-Periodic Power Law Singularity (LPPLS) model combined with a Lagrange-regularized calibration approach, following the methodology introduced by Sornette et al.~\cite{Demos2019} and Zhou \& Sornette~\cite{Demos2017}.

\subsubsection{LPPLS Model Framework}

The LPPLS model captures the accelerating price dynamics during bubbles. 
The empirical range for the log-periodic frequency parameter \( \omega \) is typically set within \([6, 13]\), based on numerous studies of financial bubbles \cite{Sornette2010},\cite{Sornette20030}. However, in practice, broader ranges such as \([4, 15]\) are often explored, especially when market dynamics are unusual or when the visibility of oscillations is affected by noise or structural shifts in the market \cite{Sornette2010}.

\subsubsection{Window Scanning and Lagrange Regularization}

To determine the most probable start date of the 2020 bubble ($t_1$), we scanned backward over a series of candidate windows $[t_1, t_2]$ and fitted the LPPLS model to the log-price series within each window. For each fit, we computed the normalized sum of squared residuals:

\[
\chi^2_{np}(t_1) = \frac{1}{N - k} \sum_{i=1}^N \left( \log p(t_i) - \hat{f}(t_i) \right)^2
\]

where $k$ is the number of free parameters in the model, and $\hat{f}(t)$ is the fitted LPPLS function.

To correct for the tendency of short windows to overfit, we applied a Lagrange regularization term:

\[
\chi^2_{\lambda}(t_1) = \chi^2_{np}(t_1) - \lambda (t_2 - t_1)
\]
The optimal bubble start time \( t_1^* \) is selected as the window start that minimizes \( \chi^2_{\lambda}(t_1) \).
The regularization coefficient $\lambda$ was estimated as the negative slope of a linear regression of $\chi^2_{np}(t_1)$ versus window size. The optimal bubble start time $t_1^*$ was defined as the window start that minimized $\chi^2_{\lambda}(t_1)$.

LPPLS fits were calculated using a grid-based search over nonlinear parameters $(t_c, \beta, \omega)$ and linear regression for $(A, B, C_1, C_2)$. This semi-parametric method provided robustness to local minima and instability.

\subsubsection{Robustness Test by Shifting $t_2$}

To verify the stability of the estimated bubble inception time, we repeated the scanning process by shifting $t_2$ backward in increments of 15 to 60 trading days. For each new $t_2$, we recalculated the Lagrange-regularized residual curve $\chi^2_{\lambda}(t_1)$.

The resulting curves were overlaid to visually assess whether the estimated bubble start remained consistent.

\subsubsection{Visualization}

Two key figures support the detection of the 2020 bubble:

\begin{itemize}
    \item \textbf{Figure 4: Bubble Start Detection (2020 Crash).} This plot shows the evolution of $\chi^2_{np}(t_1)$ and $\chi^2_{\lambda}(t_1)$ across candidate windows. The minimum of the regularized curve identifies the optimal start time.

   \textbf{Interpretation:} Although $\chi^2_{np}$ tends to favor short windows—resulting in lower residual error but potential overfitting—$\chi^2_{\lambda}$ penalizes such windows and identifies the interval between \textbf{2019-12-01} and \textbf{2020-01-05} as the most plausible inception period for the bubble. This range corresponds to the onset of super-exponential growth and log-periodic oscillations that characterize the build-up to the 2020 crash. As previously claimed, Figure~\ref{fig:3} (Plot 4) represents the optimal choice, and this section provides empirical support for that claim. Moreover, if we fix the bubble start at 2020-01-05, the estimated value of $\beta$ is approximately 0.5. Therefore, the $\beta$ parameter for \textbf{the optimally chosen window of the 2020 crash lies in the range of approximately \textbf{0.45} to \textbf{0.5}.}

    \item \textbf{Figure 5: Robustness Test (2020 Crash).} This figure overlays the $\chi^2_{\lambda}(t_1)$ curves for multiple shifted $t_2$ values. Despite the changes in end-date, all curves reach a minimum in the narrow range around early 2020.

    \textbf{Interpretation:} The clustering of minima near 2020-01 affirms the robustness of the Lagrange-regularized LPPLS approach. This indicates that the bubble start estimation is stable and does not significantly depend on the precise crash date definition.
\end{itemize}

\begin{figure}[h]
    \centering
    \includegraphics[width=.9\linewidth]{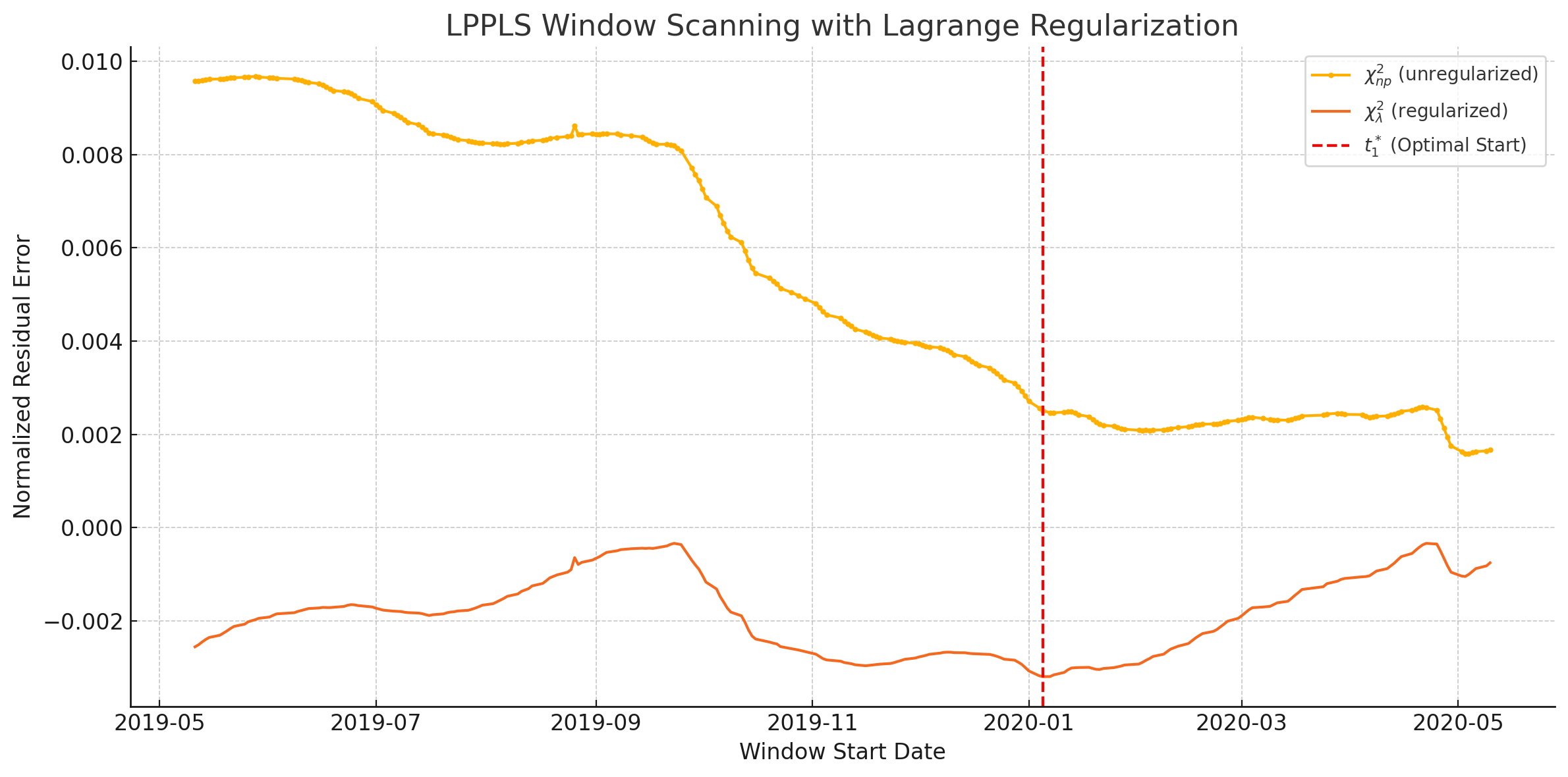}
    \caption{Bubble Start Detection Using Lagrange-Regularized LPPLS}
    \label{fig:7}
\end{figure}

\begin{figure}[H]
    \centering
    \includegraphics[width=.9\linewidth]{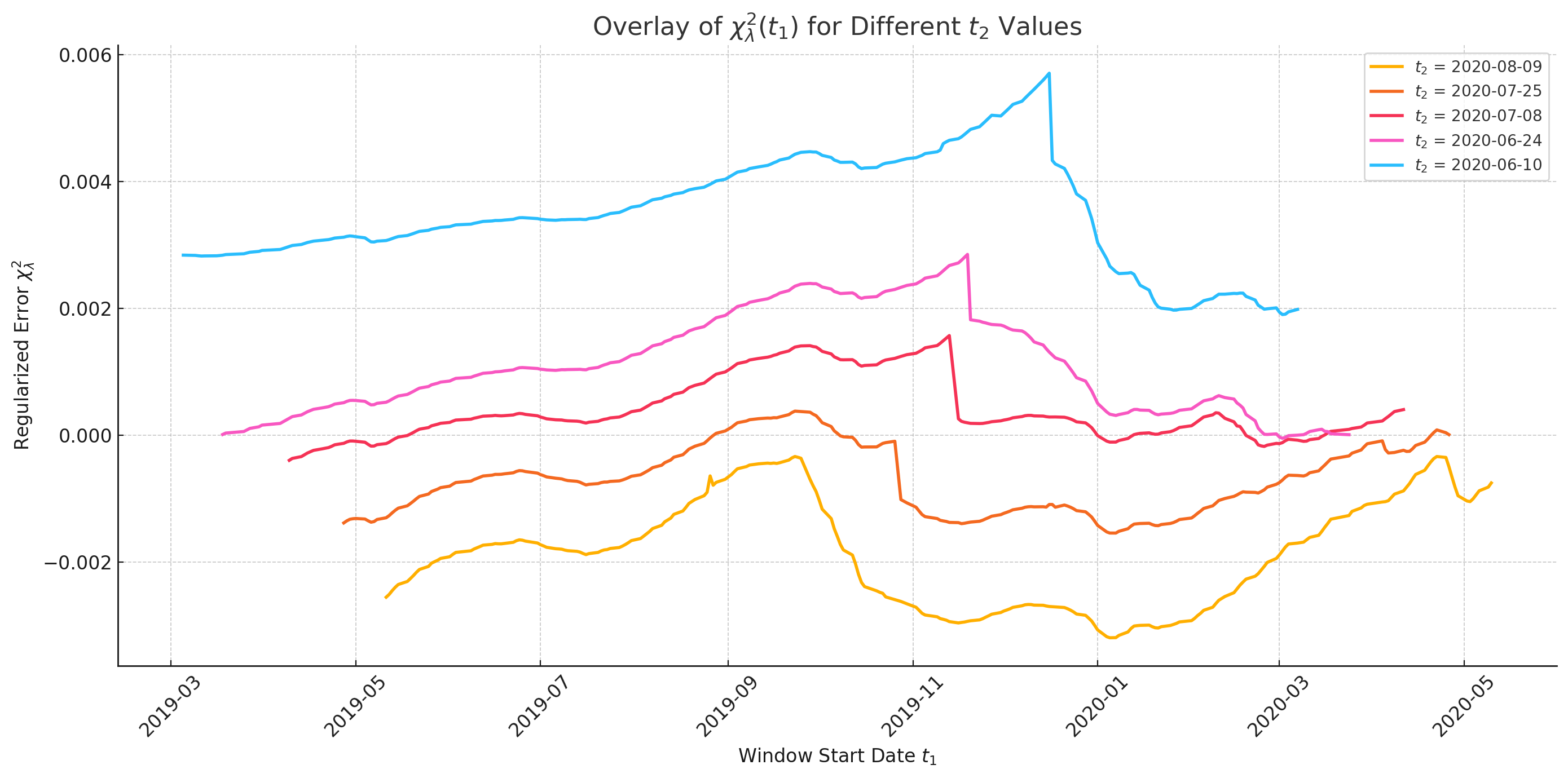}
    \caption{Robustness Test Of Bubble Start Detection (Varying $T_2$)}
    \label{fig:8}
\end{figure}

\subsection{Non-Parametric Validation of Log-Periodicity via Lomb Periodogram Analysis}

To rigorously validate the presence of log-periodic structures in the price dynamics prior to the 2023 correction phase, we employed the non-parametric spectral methodology proposed by Zhou and Sornette~\cite{ZhouSornette2002,ZhouSornette2003}. While the LPPLS model provides a parametric framework to fit financial bubbles, its sensitivity to parameter initialization and the potential for overfitting under noisy conditions necessitate an independent test of the log-periodicity hypothesis. The Lomb--Scargle periodogram offers a model-free approach to detect periodic structures in unevenly spaced or transformed time series, particularly suited to the LPPLS framework where oscillations occur in logarithmic time.

\subsection*{Methodology}

Let \( p(t) \) denote the observed price series and \( t_c \) the estimated critical time. The first step involves transforming time into logarithmic space:
\[
x = \log(t_c - t), \quad \text{for } t < t_c.
\]
Next, we fit and subtract the power-law trend from the log-price:
\[
\log p(t) \approx A + B (t_c - t)^\beta,
\]
yielding the residuals:
\[
r(t) = \log p(t) - [A + B (t_c - t)^\beta],
\]
which are expected to contain the log-periodic oscillations:
\[
r(t) \approx C (t_c - t)^\beta \cos[\omega \log(t_c - t) + \phi].
\]

To test for periodicity in \( x = \log(t_c - t) \), we compute the Lomb--Scargle periodogram \( P(\omega) \) of the residuals:
\[
P(\omega) = \frac{1}{2} \left\{ \frac{\left[ \sum_i r_i \cos \omega(x_i - \tau) \right]^2}{\sum_i \cos^2 \omega(x_i - \tau)} + \frac{\left[ \sum_i r_i \sin \omega(x_i - \tau) \right]^2}{\sum_i \sin^2 \omega(x_i - \tau)} \right\},
\]
where \( \tau \) is a phase offset that ensures time-shift invariance.

We then calculate the peak power and its corresponding frequency \( \omega^\ast \), and compare this value against surrogate data generated from four null models:
\begin{itemize}
  \item (i) White Gaussian noise,
  \item (ii) AR(1) short-memory processes,
  \item (iii) Fractional Gaussian noise (fGn) with long memory (\( H = 0.7 \)),
  \item (iv) Lévy-stable noise with heavy tails (\( \alpha = 1.7 \)).
\end{itemize}

The empirical p-value is computed as:
\[
p = \frac{1}{N} \sum_{j=1}^N \mathbb{I}(P_j^{\text{surrogate}} \geq P^\ast_{\text{real}}),
\]
where \( \mathbb{I}(\cdot) \) is the indicator function and \( N \) is the number of surrogate realizations.

\subsection*{Results for Two Detected Bubbles}

We applied this procedure to both of the bubble windows identified in Section~\ref{sec:lppls_results}: 
\begin{enumerate}
    \item \textbf{Bubble 1: Jan 20, 2019 – Aug 9, 2020}
    \item \textbf{Bubble 2: Aug 1, 2022 – May 6, 2023}
\end{enumerate}

In addition to the Lomb periodogram, we also applied the \textbf{\((H, q)\)-derivative} technique~\cite{ZhouSornette2003} to both intervals. This operator enhances weak log-periodic signals by computing a local difference operator over logarithmic time, defined as:
\[
D^{(H,q)}_t[\log p(t)] = \frac{\log p(t) - \log p(q t)}{[(1 - q)t]^H}.
\]
Log-periodic oscillations manifest as regular waves in a plot of \( D^{(H,q)}_t[\log p(t)] \) versus \( \log(t_c - t) \).

\paragraph{Bubble 1 (2019--2020):}  
The \((H, q)\)-derivative exhibits mild and less pronounced oscillatory behavior, indicating weaker---but present---log-periodic structure. The Lomb periodogram reveals a peak at \( \omega \approx 1.28 \) with spectral power of 0.062.

To assess the statistical significance of this log-periodicity, we generated 100 surrogate time series under each null model and computed the maximum spectral power. The observed peak was found statistically significant against white noise and fractional Gaussian noise (p = 0.01 and p = 0.02, respectively), but not against AR(1) (p = 0.27) or Lévy noise (p = 0.19). These results suggest that while the log-periodic signal in this bubble is weaker, it is unlikely to result from classical uncorrelated or long-memory noise alone, though autocorrelation and heavy-tailed effects remain plausible sources.

\paragraph{Bubble 2 (2022–2023):}\footnote{In the next section, we will analyze this bubble completely.}  
The \((H, q)\)-derivative exhibits sharper and cleaner oscillatory structure, suggesting stronger log-periodicity. The Lomb periodogram reveals a high peak at \( \omega \approx 1.19 \) with spectral power of 0.144.

To evaluate statistical significance, we generated 100 surrogate time series under each null model and computed the maximum spectral power. The observed peak was found statistically significant against white noise and fGn (p = 0.00 for both), but not against AR(1) (p = 0.13) or Lévy noise (p = 0.16). This supports the hypothesis that the oscillations are not random artifacts of classical or long-memory Gaussian noise, but could still emerge from more complex dynamics.

\begin{figure}[H]
    \centering
    \includegraphics[width=\textwidth]{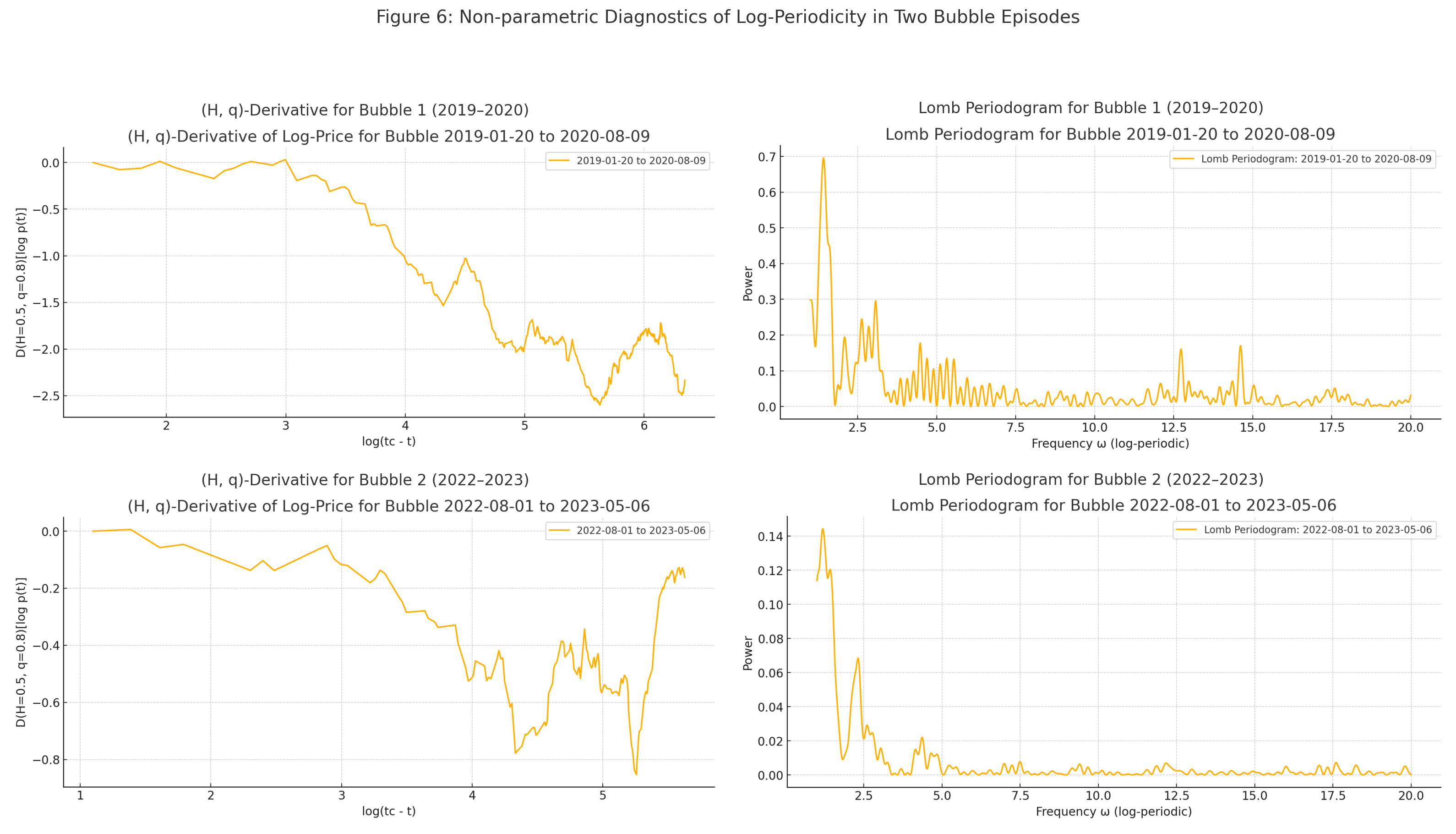}
    \caption{Non-parametric diagnostics of log-periodicity in two bubble episodes. 
    The first row shows the $(H, q)$-derivative and Lomb periodogram for Bubble 1 (2019--2020), 
    while the second row presents the same diagnostics for Bubble 2 (2022--2023). 
    The $(H, q)$-derivative detects oscillatory structure in log-space, and the Lomb periodogram reveals the spectral power 
    at log-periodic frequencies. Bubble 2 shows cleaner oscillations and stronger spectral peak compared to Bubble 1.}
    \label{fig:logperiodic_diagnostics}
\end{figure}

\paragraph{}
Figure~\ref{fig:logperiodic_diagnostics} presents the four non-parametric diagnostic plots in a two-row format. The first row shows the \((H, q)\)-derivative and Lomb periodogram for Bubble 1, and the second row for Bubble 2. These visualizations strengthen the LPPLS results by independently confirming the presence of log-periodic oscillations.

Together, \textbf{these results validate our LPPLS fitting approach and confirm the existence of critical log-periodic behavior in both historical bubbles of the Tehran Stock Exchange.}

\section{Analysis of the 2023 Tehran Stock Market Bubble}

\subsection{Overview of the 2023 Bubble}

In addition to the 2020 crash, the Tehran Stock Exchange experienced another significant
bubble that burst in July 2023. From its local peak, the Tehran Price Index (TEPIX) lost
approximately 23\% of its value, marking one of the most substantial short-term drawdowns
in the market's history. This section extends our analysis by applying the LPPLS model to
this recent bubble, further exploring the model's predictive capabilities in the context of the
Tehran stock market.

\subsection{Bubble Genesis: The “Stories” Behind the 2023 Bubble} 

The 2023 bubble emerged in a markedly different, yet equally turbulent, environment. Following the prolonged downturn after the 2020 collapse, investor sentiment began to shift in mid-2022, fueled by multiple “optimism triggers”: the lifting of price ceilings by the Securities and Exchange Organization (SEO), aggressive state messaging around a Tehran Stock Exchange (TSE) revival, and falling interest rates. 

Combined with persistent inflation and a lack of viable investment alternatives, these developments reignited speculative behavior. Critically, the dominant “story” in this bubble centered on hopeful narratives—perceived stabilization of geopolitical tensions and a post-COVID economic recovery.

As a result, the TEDPIX index rebounded sharply in late 2022, gaining more than 50\% in less than a year before culminating in a crash in May 2023. Despite improvements in market infrastructure and increased digital access, the structural fragilities of the Iranian market persisted: lack of foreign institutional capital, vulnerability to insider manipulation, and an overreliance on retail liquidity.

As with the 2020 bubble, the market once again became disconnected from economic fundamentals, demonstrating a recurring pattern of hope-driven rallies followed by abrupt corrections. In this environment, macro and policy signals appear to have coordinated investors into a self-reinforcing buying phase, aligning with the LPPLS signatures of accelerating growth and log-periodic oscillations documented for the 2023 bubble window.

\subsection{Data Collection and Preprocessing}

The data for this analysis includes Adjusted Close prices of the TEPIX from September 2022 to June 2023. Similar to the previous analysis, the data was carefully preprocessed to account for non-trading days and ensure continuity. Multiple time windows were selected to assess the model's performance over different periods leading up to the 2023 crash.

\subsection{LPPLS Model Fitting and Results}
\begin{figure}[h]
    \centering
    \includegraphics[width=1\linewidth]{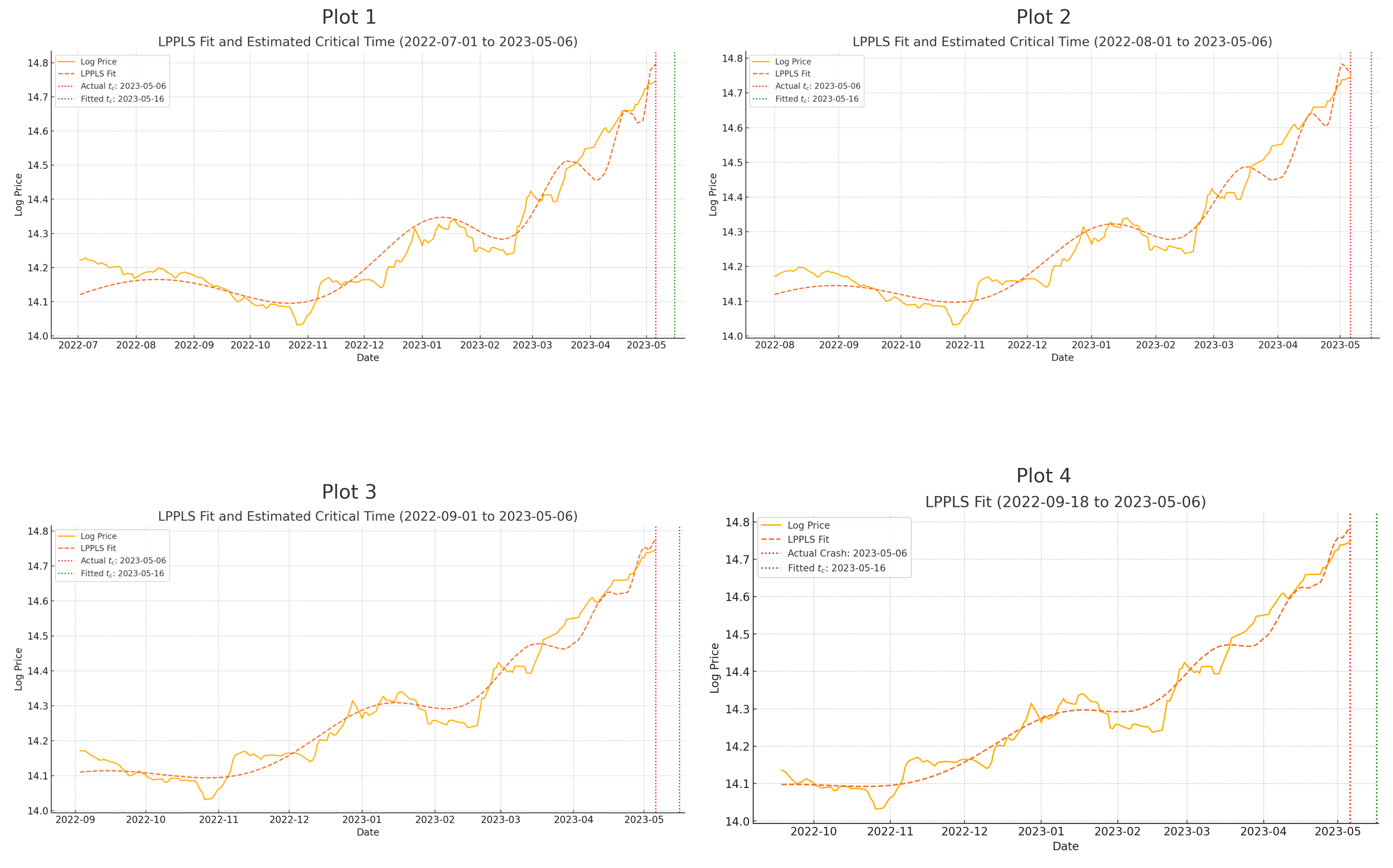}
    \caption{Results for LPPLS Model on different time frames (2023 bubble)}
    \label{fig:11}
\end{figure}

\begin{figure}[h]
    \centering
    \includegraphics[width=1\linewidth]{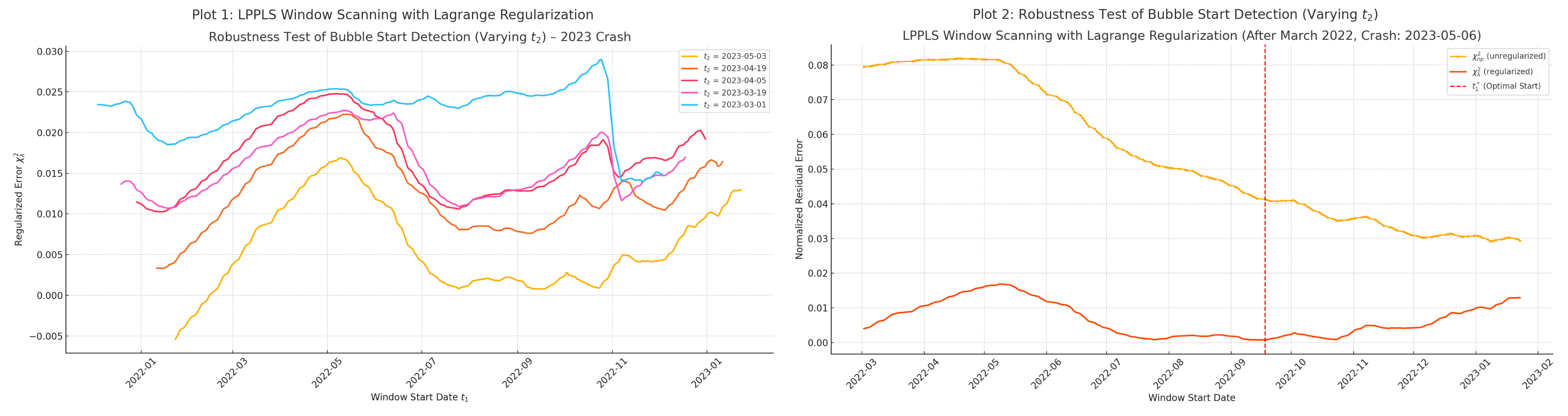}
    \caption{LPPLS window scanning and Lagrange-regularized bubble-start detection for the
2023 TSE bubble and crash.}

    \label{fig:12}
\end{figure}

We now apply the same procedure used in the previous section to the 2023 Tehran Stock Market bubble. Therefore, we do not go through the detailed steps again and instead focus on presenting the results. Figure~\ref{fig:11} shows different fits of the model. Figure~\ref{fig:12} demonstrates the stability of our fitting procedure and suggests that the
approximate inception date of the 2023 bubble lies between \textbf{2022-11-01} and
\textbf{2023-01-05}.

Note that Plot 4 in Figure~\ref{fig:11} is one of the optimal choices based on our analysis, and the\textbf{ $\beta$ value for the optimal interval of the 2023 bubble is approximately 0.2.}
In the following discussion and results, we will use this optimal $\beta$ value along with the corresponding optimal value detected for the 2020 crash.

\begin{table}[H]
\centering
\caption{LPPLS Model Parameters for Plots 1–4 (2023 Bubble)}
\begin{tabular}{lcccc}
\toprule
\textbf{Parameter} & \textbf{Plot 1} & \textbf{Plot 2} & \textbf{Plot 3} & \textbf{Plot 4} \\
\midrule
$A$       & 16.5308  & 16.6603  & 15.6006  & 15.7300 \\
$B$       & -1.3889  & -1.4758  & -0.5088  & -0.5890 \\
$C$       & 0.0444   & 0.0369   & 0.0175   & 0.0147 \\
$\beta$   & 0.1      & 0.1      & 0.2      & 0.189 \\
$\omega$  & 7.8421   & 8.5789   & 8.9474   & 9.11 \\
$\phi$    & -0.5240  & 2.1154   & 0.4143   & 3.02 \\
$t_c$ (estimated) & 2023-05-16 & 2023-05-16 & 2023-05-16 & 2023-05-22 \\
$t_c$ (days from start) & 318.0 & 288.0 & 255.0 & 240.0 \\
\bottomrule
\end{tabular}
\label{tab:lppls_plots_2023}
\end{table}

\section{Interpretation and Analysis}

The LPPLS model parameters extracted from the 2020 and 2023 bubbles in the Iranian stock market offer a basis for interpreting the speculative dynamics of each episode in light of comparable historical crashes in global markets.

\subsection{Growth Exponent \texorpdfstring{$\beta$}{β}}

The critical exponent $\beta$ characterizes the acceleration of price growth preceding a financial crash. In the 2020 Iranian bubble, our LPPLS model estimates $\beta \approx 0.45$–$0.50$, indicating a pronounced super-exponential increase in prices. This range aligns with those observed in mature financial markets—such as the 1929 DJIA crash ($\beta \approx 0.44$) and the Nasdaq bubble of 2000 ($\beta \approx 0.62$)~\cite{Johansen1999,Sornette2003}.

By contrast, the 2023 Iranian bubble yields a significantly lower exponent of $\beta \approx 0.20$, suggesting a more gradual escalation. Although such low values are rare, they have been documented in specific episodes, including the late-1990s Amazon bubble~\cite{MajumderDas2024} and the 2015 Chinese equity boom~\cite{ChenSornette2019}. This divergence in $\beta$ values highlights the variability in bubble dynamics across different market regimes.

\subsection{Phase Parameter $\phi$ and Characteristic Time Scale}

In the LPPLS framework, the phase parameter \( \phi \) traditionally modulates the alignment of log-periodic oscillations along the time axis. However, following the interpretation proposed by Sornette, it is more meaningful to express \( \phi \) in terms of a characteristic time scale \( \tau \), defined by the transformation:
\[
\phi = -\omega \log \tau \quad \Rightarrow \quad \tau = e^{-\phi / \omega}.
\]

This scale \( \tau \) represents the relative time to the critical point \( t_c \) at which the log-periodic oscillations effectively begin. In other words, it captures the onset of herding behavior and the emergence of oscillatory precursors during the bubble's growth phase. A smaller \( \tau \) indicates an earlier emergence of oscillations, whereas values of \( \tau \) close to 1 imply a more abrupt and explosive buildup immediately before the critical transition.

For the 2020 bubble (Plot 4), we estimate \( \tau \approx 0.845 \), indicating that the log-periodic structure became prominent around 15\% before the crash date. For the 2023 bubble (Plot 3), we obtain \( \tau \approx 0.955 \), suggesting the oscillatory dynamics emerged much later, within just 4.5\% of the time interval to the crash. These values are consistent with visual observations: the 2023 event exhibited a sharper and more compressed instability phase compared to the more gradually evolving 2020 bubble. Framing \( \phi \) in terms of \( \tau \) thus provides a physically interpretable and comparative measure of the temporal localization of log-periodic behavior across distinct bubbles.

\subsection{Summary of Dynamics}

The 2020 bubble represents a high-intensity, high-oscillation regime, characterized by a sharp acceleration in prices ($\beta \approx 0.46$) and a dense pattern of log-periodic oscillations ($\omega \approx 9.89$). These features suggest a speculative boom driven by strong herding dynamics, possibly amplified by macroeconomic shocks such as currency devaluation or inflationary expectations.

By contrast, the 2023 bubble unfolded more gradually, with a lower growth exponent ($\beta \approx 0.20$) but still a pronounced log-periodic structure ($\omega \approx 8.95$). This behavior is consistent with a slower build-up of speculative pressure, possibly reflecting a more informed or cautious investor base, or greater regulatory or informational feedback mechanisms.

Together, these cases demonstrate that financial bubbles—regardless of market maturity or geographical context—exhibit endogenous signatures that conform to the LPPLS framework. The variation in dynamic parameters between the two episodes further underscores the model’s flexibility in capturing different modes of bubble formation and collapse.

\section{Discussion}

\subsection*{Universality of Bubble Dynamics in the Iranian Market}

The present analysis offers compelling evidence that the Iranian stock market, despite its
economic and geopolitical isolation, exhibits speculative bubble dynamics that align closely
with those of major global financial markets. Using the Log-Periodic Power Law Singularity
(LPPLS) model, we identified two significant bubbles in the Iranian market (2020 and 2023),
with critical exponents estimated around $\beta \in [0.45, 0.50]$ and $\beta \approx 0.20$,
respectively. As summarised in Sornette's catalogue of historical bubbles~\cite{Johansen1999,
Sornette2003, Sornette2014}, LPPL fits of major crashes such as the 1929 DJIA episode,
the Nasdaq dot-com bubble of 2000, and the 2007 Chinese stock market bubble typically yield
$\beta$ values in the range $0.2$--$0.4$ and angular log-frequencies $\omega$ in the range
$6$--$9$. Our estimates for the Tehran Stock Exchange therefore lie well within, or at the
lower edge of, this empirical window.

The qualitative narratives developed in Sections 4.6 and 6.2 highlight the macroeconomic and policy environment in which the 2020 and 2023 bubbles unfolded. In 2020, renewed U.S. sanctions, currency devaluation, and state-led privatization campaigns—framed as a national wealth-building strategy—encouraged households to shift savings into equities. In 2023, a different configuration of “optimism triggers,” including relaxed price ceilings, state messaging about a market revival, and falling interest rates, again drew retail investors into the market. These episodes suggest that macro and policy shocks acted primarily as catalysts that coordinated investor expectations and amplified speculative flows. The endogenous positive-feedback mechanisms captured by the LPPLS model—herding, imitation, and reflexivity—then governed the build-up of instability and the eventual critical transition.

These findings extend existing universality tests in the LPPLS literature by adding evidence
from a politically and financially isolated market, a setting that has been largely absent from
prior empirical studies. In the context of econophysics, our results align with universal
patterns and scaling ideas discussed by Farmer, Shubik, and Smith~\cite{Farmer2005},
Yakovenko and Rosser~\cite{Yakovenko2009}, and Gabaix et al.~\cite{Gabaix2003}, showing
that endogenous feedback mechanisms and heterogeneous-agent interactions can generate
similar critical behaviors even under restricted institutional and geopolitical conditions.
This interpretation is further consistent with Bouchaud's self-organized criticality perspective
on economic and financial systems, in which fat tails, clustered volatility, and crisis avalanches
emerge from dynamics operating near critical points~\cite{BouchaudSOC}.

\begin{table}[h!]
\centering
\caption{Comparison of LPPLS parameters ($\beta$, $\omega$) for major historical bubbles
and the two Tehran Stock Exchange bubbles analyzed in this study. Values for non-Iranian
markets are taken from the LPPLS literature (Sornette, 2003; Filimonov \& Sornette, 2013;
Zhou \& Sornette, 2003).}
\label{tab:universality}
\begin{tabular}{lcc}
\hline
\textbf{Market / Bubble Episode} & \textbf{Estimated $\beta$} & \textbf{Estimated $\omega$} \\
\hline
DJIA 1929 crash               & $0.33 \pm 0.05$        & $7.0 \pm 0.5$ \\
Nasdaq 2000 dot-com bubble    & $0.34 \pm 0.04$        & $8.9 \pm 0.7$ \\
Shanghai Composite 2007       & $0.28 \pm 0.06$        & $8.8 \pm 0.6$ \\
Shanghai Composite 2015       & $0.21 \pm 0.04$        & $7.9 \pm 0.4$ \\
Bitcoin 2017 bubble           & $0.32 \pm 0.06$        & $8.3 \pm 0.5$ \\
\hline
\textbf{TSE 2020 bubble (this study)} & $\beta \in [0.45,\, 0.50]$ & $9.9$ \\
\textbf{TSE 2023 bubble (this study)} & $\beta \approx 0.20$       & $8.9$ \\
\hline
\end{tabular}
\end{table}

Our estimated parameters for the Tehran Stock Exchange bubbles fall within the broad range
documented for well-known global bubbles such as the 1929 crash, the Nasdaq 2000 episode,
and the 2015 Chinese market correction. To illustrate this universality more explicitly,
Table~\ref{tab:universality} compares the critical exponents $\beta$ and $\omega$ across
several major bubble episodes. In addition, the peak-to-trough drawdowns of the TSE 2020
and 2023 crashes (approximately $42\%$ and $23\%$, respectively) are comparable to the
30--50\% losses typically reported for major historical crashes in developed and emerging
markets~\cite{Sornette2003}.

A broader synthesis of the ranges of $\beta$, $\omega$, and crash amplitudes across the
historical bubbles catalogued by Sornette~\cite{Sornette2003}, together with the two TSE
episodes analyzed here, is provided in Appendix~II (Table~\ref{tab:meta_universality}).

Interestingly, the $\beta \approx 0.20$ observed in the 2023 bubble corresponds to a particularly
sharp super-exponential acceleration toward the critical point. While such low values are
relatively rare, similar results have been reported in analyses of the Amazon stock bubble of
the late 1990s~\cite{MajumderDas2024}, the Chinese stock market crash of
2015~\cite{ChenSornette2019}, and the South African equities surge of
2003--2006~\cite{ZhouSornette2007}. These cases demonstrate that sharply accelerating
bubbles with $\beta$ values near 0.2 can arise even in very different economic contexts,
reinforcing the hypothesis of universality in bubble dynamics.

This finding suggests that the dynamics of financial bubbles—characterized by accelerating prices, log-periodic oscillations, and eventual crashes—may reflect universal properties of financial markets, irrespective of their degree of openness, regulation, or integration into global capital flows. While Didier Sornette and collaborators have emphasized that the LPPLS exponent $\beta$ should not be interpreted as a universal constant, its clustering within a bounded range across diverse market settings underscores the presence of common underlying mechanisms\footnote{In statistical physics, a \emph{universal critical exponent} characterizes the behavior of observables (such as magnetization, correlation length, or susceptibility) near a critical point. These exponents are called \emph{universal} because they depend only on general features like the system's dimensionality and symmetry—not on microscopic details. For an accessible introduction, see~\cite{Goldenfeld1992, Cardy1996, Stanley1999}.}. In the strict sense of statistical physics~\cite{Sornette2014}, the recurring appearance of $\beta$ in a narrow range across diverse markets supports its interpretation as a robust, emergent property of collective financial behavior.

From the standpoint of complexity science, this result aligns with theories that treat financial markets as complex adaptive systems. As Farmer, Shubik, and Smith argue~\cite{Farmer2005}, the search for universality in economics—akin to that in physics—rests on identifying behavioral regularities that transcend institutional specifics. Our findings echo this philosophy: the Iranian market, dominated by retail investors, limited foreign access, and heavy regulatory influence, still converges to LPPLS dynamics observed in freer, more liquid markets. This implies that endogenous dynamics, such as herding, imitation, and reflexivity, are sufficient to produce bubble-like behaviors consistent with the LPPLS framework~\cite{Sornette2003, Farmer2005}.

Additionally, the results resonate with research by Yakovenko and Rosser~\cite{Yakovenko2009}, who demonstrate that income and wealth distributions across countries follow statistical forms analogous to those in statistical mechanics. Their work emphasizes how simple aggregate behaviors can emerge from heterogeneous agents, a principle also evident in our analysis: despite microstructural differences, the Iranian market's macro-dynamics conform to globally observed critical behaviors.

Furthermore, the power-law regularities discussed by Gabaix et al.~\cite{Gabaix2003} support the notion that large market movements—regardless of geography—stem from statistically regular agent behaviors, particularly the impact of large traders and institutions. Although the Iranian market lacks substantial foreign institutional participation, it still exhibits power-law-like accelerations and oscillatory corrections that LPPLS captures, reinforcing the idea that universal scaling laws can arise endogenously.

In summary, our findings contribute to the growing body of evidence suggesting that speculative bubbles obey generalizable principles. They offer a rare empirical example from an isolated emerging market to support the hypothesis that financial instabilities exhibit signatures of criticality, as predicted by theories of complex systems. These insights not only validate the LPPLS model's applicability across market contexts but also advance our understanding of universality in economic systems.

\section*{Acknowledgment}
I gratefully acknowledge Professor Didier Sornette for deep and constructive discussions and for his thoughtful reading of the manuscript, which significantly improved this work. I also thank my colleagues at the Physics Department \& Center for Complex Networks and Data Science (CCNSD), Beheshti University, for their encouragement and feedback. I am furthermore grateful to Bahram Shakerin, Ali Vahedi, Amir Kargaran, and Mohammad Osoolian for their valuable comments and discussions.

\section*{Data availability}
The data used in this study were obtained from the official website of the Tehran Stock Exchange (TSE) at \url{https://www.tsetmc.com}. Due to restrictions on automated data extraction and limited accessibility from outside Iran, the full cleaned dataset used in this analysis is available from the corresponding author upon reasonable request.

\section*{Code availability}
The code used for the LPPLS calibration, window optimization, and Lomb--Scargle spectral analysis is available from the corresponding author upon reasonable request.

\section*{Competing interests}
The author declares no competing interests.

\appendix
\section*{Appendix I: Robust LPPLS Calibration Algorithm (Filimonov–Sornette Method)}
\addcontentsline{toc}{section}{Appendix I: Robust LPPLS Calibration Algorithm (Filimonov–Sornette Method)}

The LPPLS (Log-Periodic Power Law Singularity) model fitting in this study follows the robust and stable estimation scheme introduced by Filimonov and Sornette~\cite{FilimonovSornette2013}. Unlike direct nonlinear least squares, this method separates the estimation of nonlinear and linear parameters, enhancing numerical stability and convergence.

\subsection*{Step 1: Preprocess Data}
\begin{itemize}
    \item Convert dates to numerical format with \( t = 0 \) at the beginning of the window.
    \item Compute the natural logarithm of prices: \( \log P_{\text{obs}}(t) \).
\end{itemize}

\subsection*{Step 2: Define the LPPLS Model}
The LPPLS equation is rewritten to separate linear and nonlinear components:
\[
\log P_{\text{LPPLS}}(t) = A + B f(t) + C_1 f(t) \cos[\omega \log(t_c - t)] + C_2 f(t) \sin[\omega \log(t_c - t)]
\]
with \( f(t) = (t_c - t)^\beta \), and \( C_1 = C \cos \phi \), \( C_2 = -C \sin \phi \).

\subsection*{Step 3: Grid Search over Nonlinear Parameters}
We construct a dense grid over:
\begin{itemize}
    \item \( t_c \in [t_{\max} + 10, t_{\max} + 200] \)
    \item \( \beta \in [0.1, 1.0] \)
    \item \( \omega \in [6, 13] \)
\end{itemize}

\subsection*{Step 4: Linear Regression for Each Grid Point}
For each grid point \((t_c, \beta, \omega)\), compute:
\begin{itemize}
    \item \( f(t) = (t_c - t)^\beta \)
    \item \( \cos[\omega \log(t_c - t)] \), \( \sin[\omega \log(t_c - t)] \)
\end{itemize}
Then solve the linear system:
\[
\log P_{\text{obs}}(t) = X \cdot \theta
\]
where \( X \) is the design matrix:
\[
X = \left[\mathbf{1}, f(t), f(t)\cos[\omega \log(t_c - t)], f(t)\sin[\omega \log(t_c - t)]\right]
\]
and \( \theta = [A, B, C_1, C_2]^T \) is estimated using Ordinary Least Squares.

\subsection*{Step 5: Select the Best Fit}
\begin{itemize}
    \item Compute the RMSE for each parameter set:
    \[
    \text{RMSE} = \sqrt{ \frac{1}{N} \sum_{i} \left( \log P_{\text{obs}}(t_i) - \log P_{\text{LPPLS}}(t_i) \right)^2 }
    \]
    \item Choose the parameter set with the lowest RMSE as the best-fit model.
\end{itemize}

\subsection*{Step 6: Recover Canonical LPPLS Parameters}
\begin{itemize}
    \item \( C = \sqrt{C_1^2 + C_2^2} \)
    \item \( \phi = \arctan2(-C_2, C_1) \)
\end{itemize}

\subsection*{Advantages of This Method}
\begin{itemize}
    \item Increased robustness to noise and poor initial guesses.
    \item Avoids instability of traditional nonlinear optimizers.
    \item Efficient and scalable across multiple detection windows.
\end{itemize}

\section*{Appendix II: Universality Ranges of LPPLS Exponents and Crash Amplitudes}
\addcontentsline{toc}{section}{Appendix II: Universality Ranges of LPPLS Exponents and Crash Amplitudes}

To place the Tehran Stock Exchange bubbles in a broader empirical context, we summarize in
Table~\ref{tab:meta_universality} the typical ranges of the LPPLS power-law exponent
$\beta$ (reported as $m_2$ in Sornette~\cite{Sornette2003}), the angular log-frequency
$\omega$, and the associated peak-to-trough crash amplitudes across several groups of
historical bubbles and antibubbles. The last two rows report the corresponding values for
the 2020 and 2023 TSE episodes documented in this study.

\begin{table}[htbp]
\centering
\caption{Summary of universality patterns across major groups of historical bubbles, as documented in Sornette (2003), together with the two Tehran Stock Exchange bubbles analyzed in this study.}
\label{tab:meta_universality}
\begin{tabularx}{\textwidth}{>{\raggedright\arraybackslash}X c c c c}
\hline
\textbf{Group / Episode Set} & \textbf{\# Episodes} & \textbf{Typical $\beta$} & \textbf{Typical $\omega$} & \textbf{Crash (\%)} \\
\hline
Major crashes (DJIA 1929, 1987, 1997, Nasdaq 2000)      
& $\sim 9$ & $0.30$--$0.60$ & $6$--$9$ & $30$--$50$ \\

Latin-American bubbles 1990s (Argentina, Brazil, Chile, Mexico, Peru, Venezuela)   
& $10$ & $0.25$--$0.70$ & $4$--$9$ & $20$--$40$ \\

1994 Western ``antibubble'' + Hong Kong
& $8$ & $0.20$--$0.50$ & $2$--$8$ & $22$--$31$ \\

Other developed/emerging bubbles (Sornette Tables 7.2 \& 8.x)
& $\sim 10$ & $0.30$--$0.70$ & $5$--$9$ & $20$--$45$ \\

\textbf{TSE 2020 (this study)}
& $1$ & $[0.45, 0.50]$ & $\approx 9.9$ & $\approx 42$ \\

\textbf{TSE 2023 (this study)}
& $1$ & $\approx 0.20$ & $\approx 8.9$ & $\approx 23$ \\
\hline
\end{tabularx}
\end{table}

\clearpage
\bibliographystyle{plain}

\end{document}